# The bixbyite framework as a platform for frustrated noncollinear magnetism: resolving the magnetic ground state of β-$Fe_2O_3$

Chenjun Tang[1], Ondřej Malina [2,3], Jiří Tuček [4], Francois Fauth[5], Martí Gich[1], and José Luis García-Muñoz[1,*]

[1] *Institut de Ciència de Materials de Barcelona (ICMAB-CSIC), Carrer dels Til.lers, 08193 Cerdanyola del Vallès, Spain.*

[2] *Regional Centre of Advanced Technologies and Materials, Czech Advanced Technology and Research Institute, Palacky University, Olomouc, Slechtitelu 27, 77900 Olomouc, Czech Republic.*

[3] *Nanotechnology Centre, Centre for Energy and Environmental Technologies, VSB-Technical University of Ostrava, 17. listopadu 2172/15, 70800, Ostrava-Poruba, Czech Republic*

[4] *Research Centre, Faculty of Electrical Engineering and Informatics, University of Pardubice, Studentská 95 53002, Pardubice, Czech Republic*

[5] *CELLS-ALBA Synchrotron, 08290 Cerdanyola del Vallès, Barcelona, Spain*

* Corresponding Author. Email: garcia.munoz@icmab.es

## Abstract

Magnetic behavior across $Fe_2O_3$ polymorphs varies widely despite identical chemistry, highlighting crystal architecture as a key determinant of exchange topology, magnetic anisotropy, and ultimately magnetic order. Here, using neutron and synchrotron X-ray diffraction, we establish the magnetic ground state of the poorly understood bixbyite β-$Fe_2O_3$ polymorph and uncover the structural origin of its strong frustration. Below the Néel temperature, a noncollinear antiferromagnetic state emerges through activation of the mH1+ irrep at the H point [**k** = (1,1,1)] and the antitranslation (1′|1/2,1/2,1/2), breaking the body centering and yielding two interpenetrating primitive cubic magnetic subcells with inverted moments and nonpolar type-IV symmetry. Under exclusively antiferromagnetic $Fe^{3+}$–O–$Fe^{3+}$ interactions, β-$Fe_2O_3$ exhibits a large frustration index ($f \approx 7.56$). This behavior originates from the intrinsic geometry of the bixbyite lattice: two magnetic sublattices with distinct point symmetries and anisotropy constraints are embedded in a three-dimensional exchange network containing interconnected triangular and hexagonal motifs. In {111} planes, Fe2 ions form hexagonal rings interconnected by frustrated triangular units, while locally Ising-like Fe1 ions occupy the ring centers. Our results thus identify the bixbyite architecture as a promising general platform for frustrated noncollinear magnetism. Extending this structural framework to other magnetic transition-metal or 4*f* ions opens a materials space for engineering competing exchange interactions and anisotropies, potentially stabilizing new noncollinear and field-tunable magnetic states.

## I. INTRODUCTION

Iron oxides are well-known multifunctional materials with many technological applications spanning fields such as catalysis, environmental remediation, biomedicine, energy storage or electronics [1–3]. The ubiquitous iron (III) oxide (ferric oxide, $Fe_2O_3$) is a paradigmatic polymorphic transition-metal oxide. Polymorphism in transition-metal oxides provides a direct route to tuning the exchange topology, magnetocrystalline anisotropy, and electronic structure without altering the chemical composition. $Fe_2O_3$ is a canonical example: several polymorphs are known, each built from a distinct arrangement of $Fe^{3+}$ cations within close-packed oxygen frameworks and therefore characterized by different Fe coordination polyhedra, site symmetries, and lattice connectivity. These structural degrees of freedom result into sharply different magnetic phase diagrams.

At ambient pressure, the best-established crystalline polymorphs are $\alpha$-$Fe_2O_3$ (hematite), $\gamma$-$Fe_2O_3$ (maghemite), $\varepsilon$-$Fe_2O_3$, and $\beta$-$Fe_2O_3$, while pressure treatment can stabilize additional forms such as $\zeta$-$Fe_2O_3$ [4–11]. The structures of the four crystalline $Fe_2O_3$ polymorphs known under ambient conditions ranges from the corundum-type $\alpha$ phase (*R-3c*) to the inverse-spinel-related $\gamma$ phase, to the non-centrosymmetric orthorhombic $\varepsilon$ phase ($Pna2_1$), and finally the bixbyite-type $\beta$ phase (cubic *Ia-3*) comprising only octahedrally coordinated $Fe^{3+}$ but split into two inequivalent crystallographic sites [4,9,12,13].

Across these polymorphs, the magnetic behavior is largely controlled by (i) the topology and metrics of Fe–O–Fe superexchange pathways, (ii) the presence/absence of multiple crystallographic Fe sublattices, including tetrahedral versus octahedral coordination, and (iii) the magnetocrystalline anisotropy set by local distortions and spin–orbit-driven antisymmetric exchange. The resulting magnetic phases include canted antiferromagnetism with spin-reorientation transitions in $\alpha$-$Fe_2O_3$ (hematite), ferrimagnetism in $\gamma$-$Fe_2O_3$ (maghemite) and $\varepsilon$-$Fe_2O_3$, and low-temperature antiferromagnetism in $\beta$-$Fe_2O_3$.

The thermodynamically stable $\alpha$-$Fe_2O_3$ polymorph undergoes the Morin transition near ~260 K, below which it behaves as an easy-axis antiferromagnet with moments oriented close to the crystallographic c axis, while above the Morin temperature the spins lie in the basal plane and acquire a weak ferromagnetic component due to canting enabled by Dzyaloshinskii–Moriya interactions [14,15]. Long-range antiferromagnetic order persists until a very high Néel temperature (~955 K), reflecting strong superexchange in the corundum lattice. In the $\alpha$ phase, magnetism is robust, and spin reorientation is controlled primarily by competing magnetic anisotropy terms whose sign changes with temperature and can be altered by size, strain, and external fields.

In contrast, $\gamma$-$Fe_2O_3$ (maghemite) instead derives from the spinel topology, with $Fe^{3+}$ occupying tetrahedral A and octahedral B sublattices and charge neutrality enforced by cation vacancies predominantly on octahedral positions [16]. Maghemite is ferrimagnetic and in the form of ultrafine particles, it has served as a model system to study superparamagnetism [4,17,18]. Thermal stability complicates direct determination of its Curie temperature, $T_C$ (estimated indirectly in the range 590-675°C), because $\gamma$-$Fe_2O_3$ tends to transform upon heating and reported $T_C$ values can vary depending on the preparation method. A crucial structural aspect is vacancy ordering, which can lower the *Fd-3m* symmetry of the disordered spinel aristotype to ordered variants such as $P4_132$ or tetragonal superstructures.

$\varepsilon$-$Fe_2O_3$ is orthorhombic and non-centrosymmetric, hosting four crystallographically inequivalent Fe sites (three octahedral and one tetrahedral environment), with pronounced site-dependent polyhedral distortions [19,20]. These structural features give rise to frustration and strong magnetocrystalline anisotropy, gigantic coercive fields (tens of kOe) and successive magnetic phase transitions that involve concomitant structural changes. $\varepsilon$-$Fe_2O_3$ remains ferrimagnetically ordered far above the previously assumed ~500 K boundary (limit of the hard ferrimagnetic regime), and a second ferrimagnetic order persists up to ~ 850 K [10]. In the soft ferrimagnetic phase, only the spins of two Fe sublattices are robustly ordered and the net ferromagnetic component is strongly reduced. A complex low-temperature transition in the ~85–150 K range is still a matter of discussion.

$\beta$-$Fe_2O_3$, which exists only in nanosized form, adopts the bixbyite-type cubic structure (space group *Ia-3*), built exclusively from $FeO_6$ octahedra but split into two nonequivalent octahedral $Fe^{3+}$ sites [5,21]. $\beta$-$Fe_2O_3$ is thermally metastable and transforms to the $\alpha$ phase at elevated temperatures. The topotactic transformation of $\beta$-$Fe_2O_3$ to $\alpha$-$Fe_2O_3$ by heat treatment was reported by Danno et al [5], underscoring that polymorph selection in $Fe_2O_3$ is frequently governed by kinetic constraints and by the ability of the oxygen framework to reorganize with minimal atomic diffusion [4,5]. Within this polymorph family, $\beta$-$Fe_2O_3$ is exceptional, as, in sharp contrast to $\alpha$-, $\gamma$-, and $\varepsilon$-$Fe_2O_3$, it is magnetically disordered (paramagnetic) at room temperature. It only develops an antiferromagnetic order well below room temperature, $T_N$ typically reported in the range of $\approx$100–120 K [4,6,7]. An early Mössbauer study established $T_N$ =106.5 ± 1.5 K [7] and later studies reported values near ~110–119 K [4,6,22].

A noncollinear spin arrangement was previously proposed by Malina *et al.* on the basis of Mössbauer measurements [3]. In addition, a field-induced spin-reorientation (spin-flop) transition was inferred from in-field Mössbauer spectra collected at 5 K under applied magnetic fields of 2–3.5 T [3]. A deviation from ideal antiferromagnetic behavior below $T_N$ was also invoked to account for a small net ferromagnetic component of 0.003 $\mu_B$ per Fe cation at 5 K derived from magnetization measurements.

A neutron diffraction study of the $Fe^{3+}$ spin arrangement in the β phase would be desirable, because this polymorph's magnetic properties remain unknown, and it is the only $Fe_2O_3$ polymorph whose magnetic ordering has yet to be established. Here, we investigate the Néel transition, the effects of geometrical magnetic frustration, and the magnetic ordering in β–$Fe_2O_3$. Beyond establishing its magnetic ground state, our results identify the bixbyite crystal architecture itself as an effective structural platform for frustrated noncollinear magnetism. As a consequence, β–$Fe_2O_3$ has been found to exhibit the highest reported frustration index among binary magnetic oxides. The combination of two crystallographically distinct magnetic sublattices with different local point symmetries and anisotropy constraints, together with the singular three-dimensional topology of exchange interactions, naturally promotes competing magnetic interactions and noncollinear spin configurations. In particular, the coexistence of hexagonal and triangular magnetic motifs, the embedding of locally anisotropic magnetic ions at the centers of hexagonal magnetic rings, and the symmetry-imposed magnetocrystalline anisotropy generate a geometrically frustrated magnetic network whose stabilization is governed by symmetry rather than by chemical complexity. These characteristics are intrinsic to the bixbyite framework, indicating that bixbyite-derived compounds containing other magnetic transition-metal or rare-earth ions provide a broad materials platform for engineering frustrated magnetic states, competing anisotropies, and potentially novel functional properties arising from complex noncollinear spin textures.

## II. SAMPLE PREPARATION AND EXPERIMENTAL METHODS

**Sample preparation.** The β-$Fe_2O_3$ sample was prepared by thermal decomposition of $Fe_2(SO_4)_3 \cdot 5H_2O$ in the presence of NaCl (see also ref. [6]). Commercial $Fe_2(SO_4)_3 \cdot 5H_2O$ was used directly without any preliminary dehydration or additional pretreatment. The iron sulfate precursor was thoroughly mixed with NaCl in a molar ratio of 3:1, and the resulting powdered mixture was placed in a furnace and annealed in air at 400 °C for 1 h. After thermal treatment, the obtained product was thoroughly washed with distilled water to remove residual soluble salts and subsequently dried, yielding β-$Fe_2O_3$.

**Experimental methods.** Neutron powder diffraction (NPD) data were collected at the high-flux reactor of the Institut Laue Langevin (ILL) in Grenoble, France. Measurements were carried out on the high-intensity D1B diffractometer ($\lambda$=1.28 Å) equipped with an Orange cryostat. NPD patterns were collected both at fixed temperatures, with an acquisition time of 30 min per pattern, and in continuous mode during temperature ramps between 10 and 300 K at a heating rate of 0.8 K/min (5 min per pattern). β-$Fe_2O_3$ was also characterized by synchrotron X-ray powder diffraction (XRPD) at the BL04-MSPD beamline [23] of the ALBA Synchrotron Light Facility (Cerdanyola del Vallès, Spain). The powder sample was loaded into 0.7-mm-diameter borosilicate glass capillary that was continuously rotated during data acquisition. A short wavelength ($\lambda$= 0.4137 Å) was selected to minimize absorption. XRPD patterns were collected at selected temperatures across the magnetic transition. For XRPD data collection, a high-angular-resolution multianalyzer detector (MAD) was used, in combination with a He-flow cryostat (Dynaflow). The macroscopic magnetic characterization was performed using a Superconducting Quantum Interferometer Device (SQUID) from Quantum Design.

For the magnetic symmetry analysis, we used the crystallographic tools from the Bilbao Crystallographic Server [24–27] and the ISOTROPY software suite [28]. Structural and magnetic refinements were performed using the Fullprof set of programs [29]. Illustrations of the crystal and magnetic structures were generated using the VESTA program [30].

## III. RESULTS AND DISCUSSION

### III-a Crystal structure of β-$Fe_2O_3$

The room-temperature neutron powder diffraction pattern was refined in the *Ia-3* space group (Fig. 1), using as the starting structural model that previously reported by Danno *et al.* [5]. The structural parameters derived from the Rietveld refinement of the neutron data at 300 K are summarized in Table I. Selected Fe-O interatomic distances and bond angles are given in Table II. β-$Fe_2O_3$ presents a body-centered cubic structure with space group symmetry *Ia-3* (No. 206) and a lattice constant of 9.4054(3) Å at 300 K. Its cubic crystal structure is of a bixbyite type. The Fe atoms occupy two Wyckoff positions of different symmetry: Fe1 is at the 8*a* [.**-3**.] site, (0,0,0); and the second iron, Fe2, at the 24*d* [**2**..] site, (x,0,1/4). There is only one oxygen site, located at the general position 48*e* (x,y,z). The two centering vectors of this space group are (0,0,0) and (1/2,1/2,1/2).

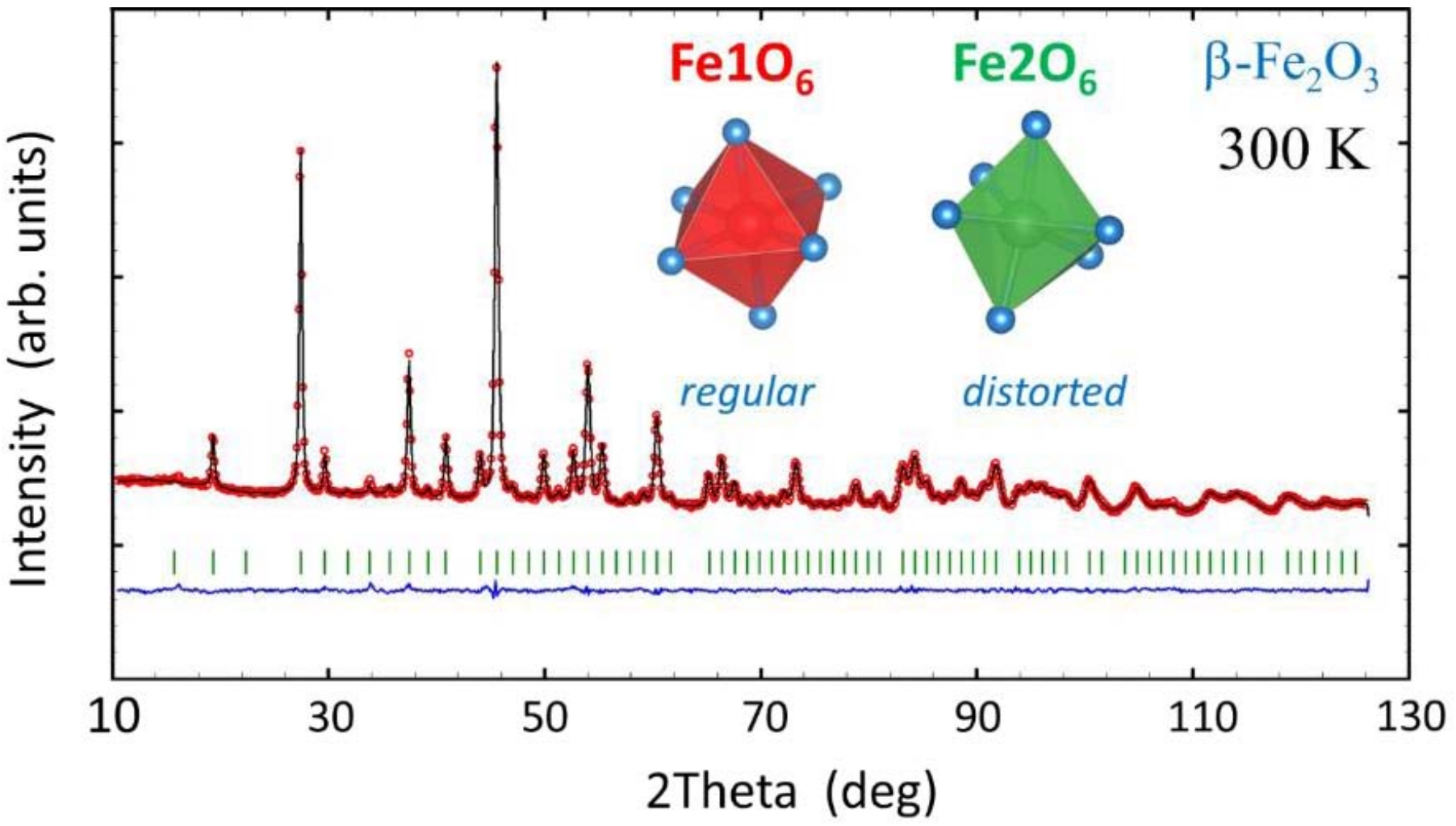

**Figure 1.** Rietveld refinement (black line) of the room temperature D1B@ILL ($\lambda$ = 1.28 Å) neutron diffraction pattern (red circles) of β-$Fe_2O_3$ at 300 K (*Ia-3*). Structural parameters are given in Table I.

**Table I**. Refined crystal structure (cell parameters, atomic coordinates and isotropic displacement parameters (*B* )) for β-$Fe_2O_3$ obtained from Rietveld refinement of neutron powder diffraction data at 300 K and 10 K (space group Ia-3, No. 206).

| **Temperature 300 K** | *a* = 9.4054(3) Å | $\chi$ 16.90 | $\mathbf{R_f}$ 1.44 | $\mathbf{R_B}$ 2.11 | |
|---|---|---|---|---|---|
| **Atom** | **x** | **y** | **z** | **(Å²)** | **WP** |
| Fe1 | 0.00000 | 0.00000 | 0.00000 | 0.153(43) | *8a* |
| Fe2 | 0.28429(10) | 0.00000 | 0.25000 | 0.264(31) | *24d* |
| O1 | 0.14269(18) | 0.13196(16) | -0.09344(17) | 0.252(24) | *48e* |
| **Temperature 10 K** | *a* = 9.3890(3) Å | $\chi$ 8.89 | $\mathbf{R_f}$ 1.50 | $\mathbf{R_B}$ 2.14 | |
| **Atom** | **x** | **y** | **z** | **(Å²)** | **WP** |
| Fe1 | 0.00000 | 0.00000 | 0.00000 | -0.097(49) | *8a* |
| Fe2 | 0.28410(12) | 0.00000 | 0.25000 | 0.110(37) | *24d* |
| O1 | 0.14295(22) | 0.13151(20) | -0.09377(21) | 0.072(29) | *48e* |

**Table II.** Selected Fe–O interatomic distances, bond angles, bond valence sums (BVS), and octahedral distortion parameters (Δ) for β-$Fe_2O_3$ obtained from Rietveld refinement of powder neutron diffraction data at 300 K and 10 K.

| **β-$Fe_2O_3$** | **300 K** | **10 K** |
|---|---|---|
| | **Distance (Å)** | **Distance (Å)** |
| Fe1−O1 (×6) | 2.0279(17) | 2.0256(19) |
| ⟨ Fe1−O1 ⟩ | 2.0279(17) | 2.0256(19) |
| BVS | 2.997(6) | 3.016(6) |
| Distortion (Δ) | 0 | 0 |
| Fe2−O1 (×2) | 1.9602(19) | 1.9587(20) |
| Fe2−O1 (×2) | 2.0442(17) | 2.0357(19) |
| Fe2−O1 (×2) | 2.0867(18) | 2.0869(20) |
| ⟨ Fe2−O1 ⟩ | 2.0304(7) | 2.0273(8) |
| BVS | 3.008(6) | 3.033(7) |
| Distortion (Δ) | $6.710\times10^{-4}$ | $6.711\times10^{-4}$ |
| | **Angle (deg.)** | **Angle (deg.)** |
| Fe1−O1−Fe2 | 99.37(8) | 99.58(8) |
| Fe1−O1−Fe2 | 126.41(8) | 126.18(9) |
| Fe1−O1−Fe2 | 97.98(7) | 97.90(8) |
| Fe2−O1−Fe2 | 126.41(7) | 126.52(8) |
| Fe2−O1−Fe2 | 98.17(8) | 98.28(9) |
| Fe2−O1−Fe2 | 100.91(9) | 100.78(10) |

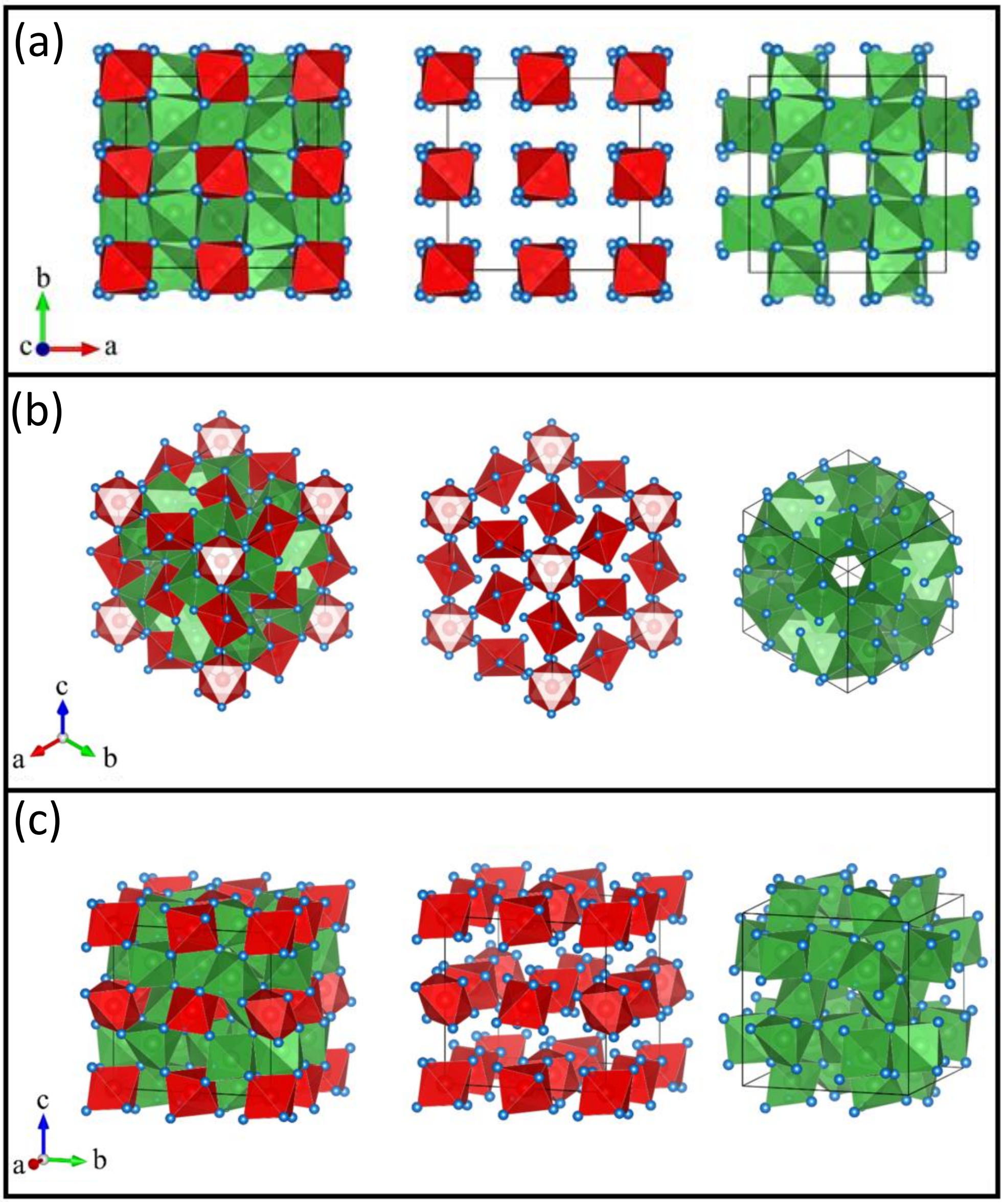

**Figure 2.** Room-temperature crystal structure of $\beta$-$Fe_2O_3$. Projections of the *Ia-3* structure refined at 300 K for different unit-cell orientations. The two crystallographically distinct Fe1 (red) and Fe2 (green) sites, together with the O atoms (blue), are shown in the figure. For each orientation, three representations are provided: (left) the complete structure, (center) only the regular $Fe1O_6$ octahedra (red), and (right) only the distorted $Fe2O_6$ octahedra (green).

Figure 2 presents the refined cubic structure in three unit-cell orientations. For each orientation there are three complementary views: the left panel shows all atoms, the center panel displays only the octahedra around Fe1, and the right panel shows only the octahedra around Fe2 to aid visualization. In these representations, the Fe1 sites are shown in red, Fe2 sites in green, and oxygen atoms in blue. The two types of $FeO_6$ octahedra, regular and distorted, are also shown as insets in Fig. 1. In the more regular $Fe1O_6$ octahedron, the six Fe1–O1 bond lengths are equivalent (2.0279(17) Å), whereas the O1–Fe1–O1 angles formed by the apical and basal oxygen atoms deviate from 90° by ±8.45° (98.45° and 81.55°, respectively). In contrast, the Fe2$O_6$ octahedron is significantly more distorted, exhibiting two elongated Fe2–O1 bonds (2.0422(17) and 2.0867(18) Å) and one shortened Fe2–O1 bond (1.9602(19) Å). In the β phase, each $FeO_6$ octahedron is connected to two neighboring $FeO_6$ octahedra by edge sharing and to two others through corner sharing.

To gain insight into the low-temperature structure and assess possible symmetry changes induced by magnetic ordering, the sample was also investigated by synchrotron XRPD using a high-angular-resolution multianalyzer detector (MAD). A refinement of the synchrotron XRPD pattern with high statistics collected at 5 K is shown in Fig. S1 [31]. No structural symmetry change was detected upon cooling, and the structure of the antiferromagnetic phase can be described within the *Ia-3* space group.

### III-b Antiferromagnetic transition in β-$Fe_2O_3$.

Using a superconducting quantum interference device magnetometer (SQUID), zero-field-cooled (ZFC) and field-cooled (FC) dc magnetic susceptibility, $\chi$, measurements were performed under an applied field of 2000 Oe, upon warming from 10 to 350 K. Figure 3(a) shows the χ(T) curves, where a pronounced maximum in χ(T) indicates the onset of antiferromagnetic ordering at $T_N$ ≈112 K, in good agreement with earlier reports [7,22,32]. As shown in Fig. 3, the ZFC and FC curves diverge below ~265 K, well above the Néel transition by more than 150 K.

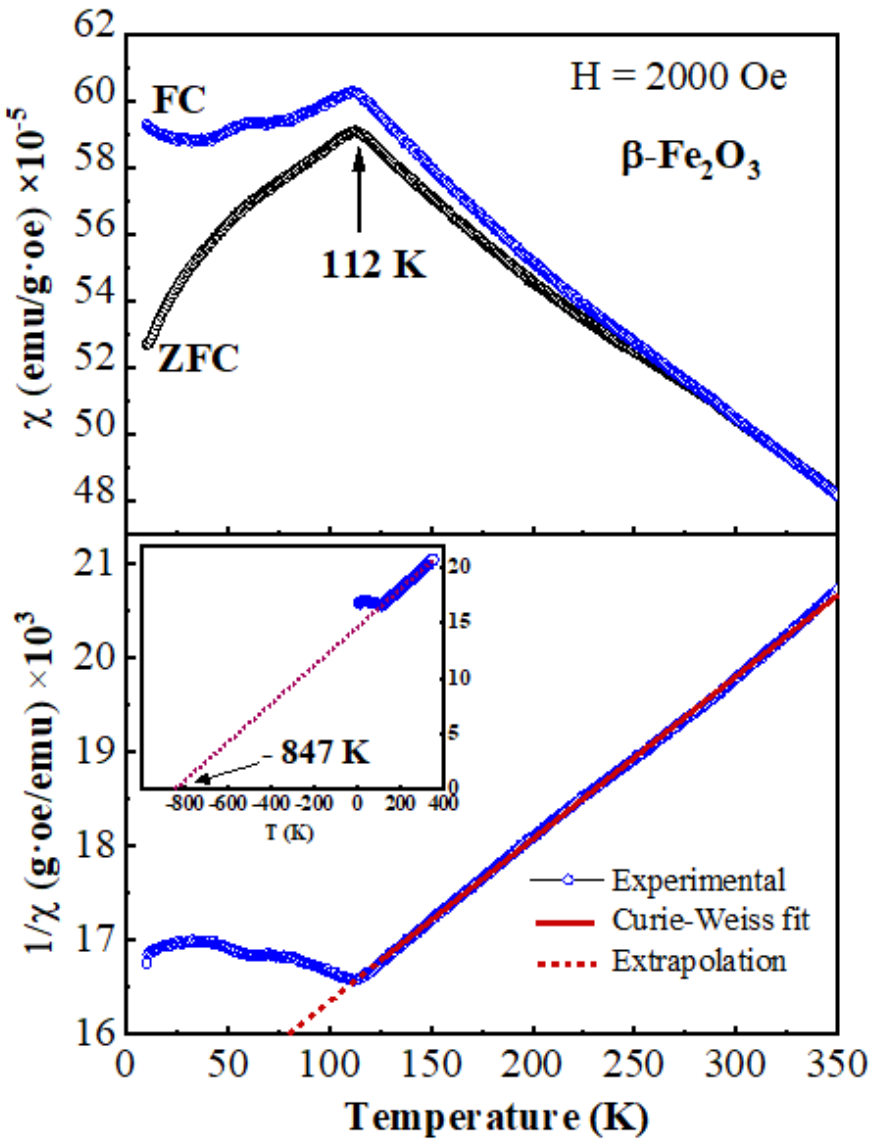

**Figure 3.** Magnetic susceptibility. (a) FC and ZFC susceptibility curves of the β-$Fe_2O_3$ sample (H=2 kOe). (b) Inverse susceptibility and Curie-Weiss fit above the antiferromagnetic transition. Inset: extrapolated Curie temperature ($\theta_C$ = – 847 K).

Fig. 3(b) shows the Curie–Weiss fit to the inverse susceptibility of our sample. The fit yields a Weiss temperature $\theta_C = -847$ K and a Curie constant $C$=9.53x10$^{-6}$ m$^3$·K·mol$^{-1}$. As previously reported in the study by Malina *et al.* ($\theta_C = -757$ K, $H$ = 1 kOe) [6], the large absolute value and negative sign of $\theta_C$ indicate strong exchange interactions, dominated by antiferromagnetic contributions. We obtained a frustration index [33] of $f = |\theta_C/T_N| \approx 7.56$. This value places this compound, to our knowledge, as the binary magnetic oxide with the highest reported frustration index, including binary oxides in which magnetic frustration originates from either geometrical effects or competing further-neighbor interactions. By way of comparison, the $f$ values for MnO, FeO, CuO or $Cr_2O_3$ are ~5, ~4, ~2 and ~1.5, respectively [34]. The origin of magnetic frustration in this iron oxide polymorph will be addressed later.

**III-c Neutron diffraction: the non-collinear antiferromagnetic order of β-$Fe_2O_3$**

The magnetic transition of β-$Fe_2O_3$ was investigated by neutron powder diffraction (NPD) at the high-flux reactor of the Institut Laue Langevin (ILL, Grenoble, France). After cooling to 10 K, neutron diffraction patterns were collected on D1B every 5 min while warming the sample to room temperature at 0.8 K/min. The 2θ-$T$ projection of the neutron diffraction intensity across the transition is shown in Fig. 4. It reveals the onset and evolution of additional reflections of magnetic origin below $T_N$. NPD patterns were also collected, with longer acquisition times, at selected fixed temperatures (10, 130, and 300 K) for structural and magnetic refinements. Fig. S2 [31] shows the difference between the neutron patterns measured at 10 K (magnetically ordered) and 130 K (paramagnetic), highlighting the main magnetic reflections appearing below the magnetic transition. The main magnetic peaks occur at positions forbidden by the cubic *Ia-3* symmetry, indicating either a nonzero propagation vector or a lowering of symmetry. The latter was ruled out in Sec. III-a and, as shown below, the magnetic reflections can be indexed with the propagation vector **k**=(1,1,1) without invoking an enlarged magnetic unit cell. The corresponding extinction conditions arise from the body centering of the paramagnetic structure, which is broken by the magnetic ordering. Fig. 5 shows the temperature evolution of the integrated intensity of the strongest magnetic reflection. No additional magnetic transitions were observed in the compound at lower temperatures.

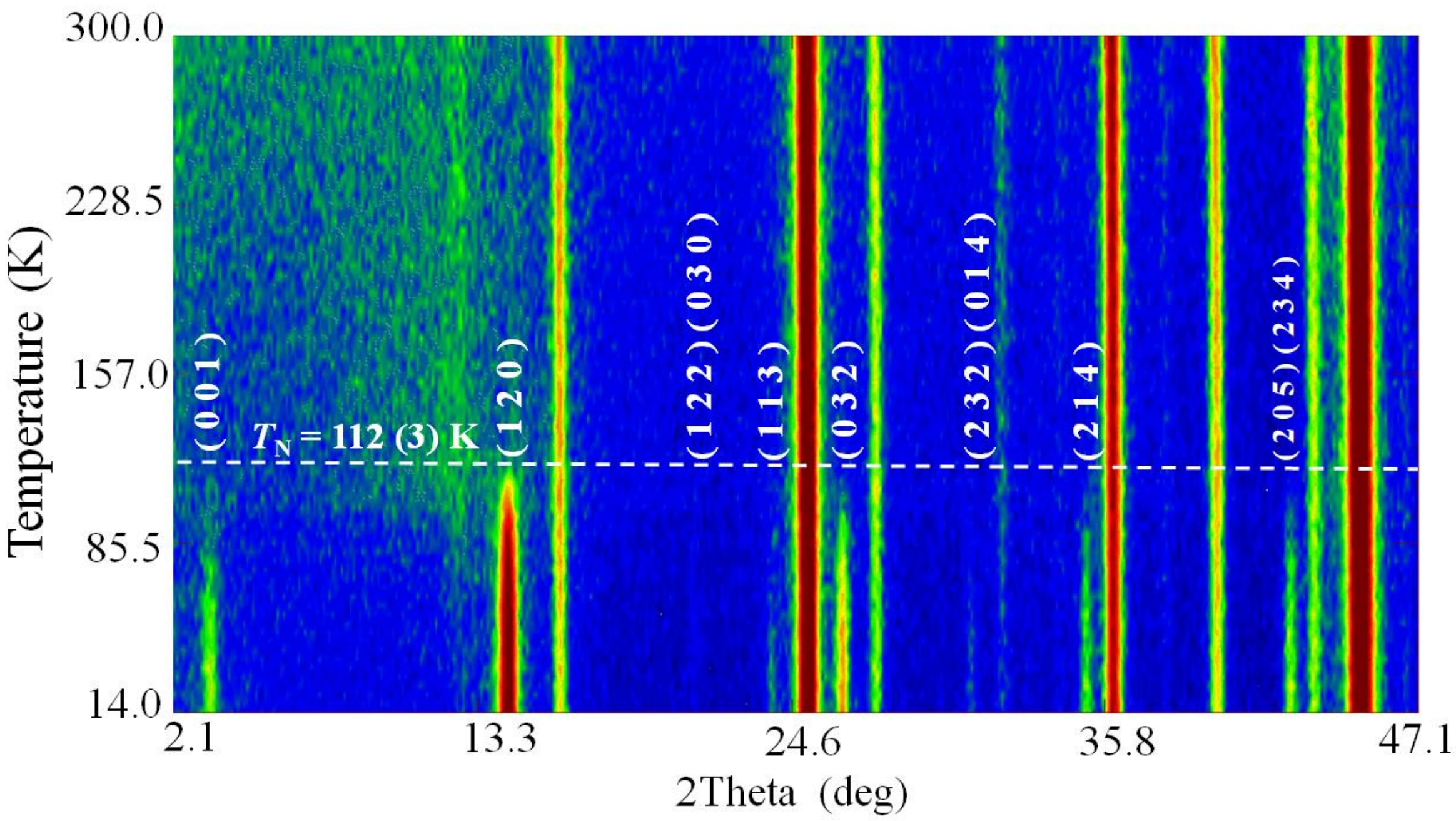

**Figure 4.** Contour maps of β-$Fe_2O_3$ showing the 2θ-$T$ projection of the temperature evolution of the neutron diffraction intensity (D1B@ILL, $\lambda$ = 1.28 Å). The low-$Q$ region highlights the onset and temperature evolution of the main magnetic reflections below the antiferromagnetic transition.

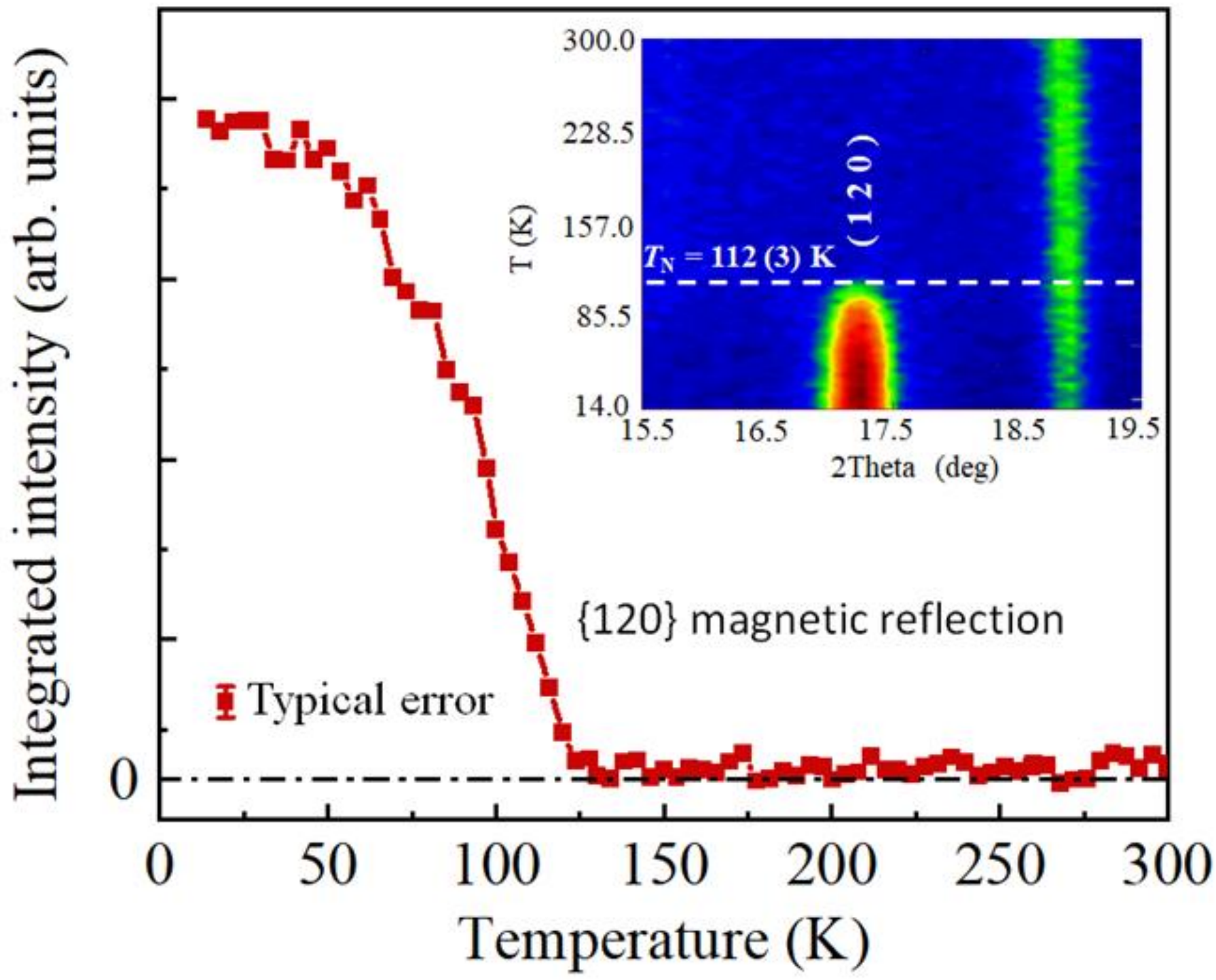

**Figure 5.** Evolution of the magnetic order. Temperature evolution of the integrated neutron intensity of the {120} magnetic reflection.

**Noncollinear antiferromagnetic ground state.** The magnetic structure of β-$Fe_2O_3$ was resolved, confirming a noncollinear spin arrangement with propagation vector **k**=(1,1,1) (H-point). Below $T_N$, the magnetic reflections and their intensities are well reproduced by the magnetic space group (MSG) $P_Ia\text{-}3$ (No. 205.36 in BNS notation [35]), which preserves the dimensions of the crystalline unit cell. This MSG is maximal subgroup of the paramagnetic space group with **k**=(1,1,1), and the parent setting (**a**,**b**,**c**;0,0,0) can be used to describe the magnetic ordering in the standard setting. We thus confirmed the loss of the translational symmetry associated with $\mathbf{t}_I$ =(1/2,1/2,1/2), i.e., the *I*-centering, in the antiferromagnetic phase. In Fig. 6 we present the full refinement of the structural and magnetic intensities recorded in the NPD pattern collected at 10 K. The magnetic structure refined at 10 K is fully described in Table III. This table exposes all the details of the magnetic ground state, including the magnetic space and point groups, and the components of refined ordered magnetic moments at the two Fe sites. The magnetic order is non-collinear, and the spins are also non-collinear within each of the two magnetic orbits of the magnetic structure. The corresponding structural parameters and refined atomic coordinates at 10 K are listed in Table I, for comparison with the structure at 300 K. The magnetic point group is *m-3.1'* (No. 29.2.110) and it is non-polar ( incompatible with spontaneous ferroelectric ordering). This point symmetry is also incompatible with macroscopic ferromagnetism [36]. Our magnetic refinements indicate that Fe1 and Fe2 magnetic sublattices are not completely polarized at 10 K (*m*=4.02(12) and 3.65(6) $\mu_B$, respectively). The reduction of the ordered moments relative to the expected values for half-filled $Fe^{3+}$ ($3d^5$) ions may arise from frustration effects and/or from the nanosized nature of the sample [6].

The antiferromagnetic structure of β-$Fe_2O_3$ is illustrated in Fig. 7. To provide a clearer view of the spin arrangement, we have represented several projections of the magnetic structure along distinct zone axes: [001], [111], and [110]. Given the complexity of this antiferromagnetic order, three representations are shown for each projection to facilitate visualization: the first (left panel) displays all magnetic atoms of the unit cell; the other two panels (center and right) show only the magnetic moments at the Fe1 (red) and Fe2 (green) sites, respectively. Owing to the cubic symmetry, the projections of the magnetic order along the [100], [010], and [001] zone axes are identical.

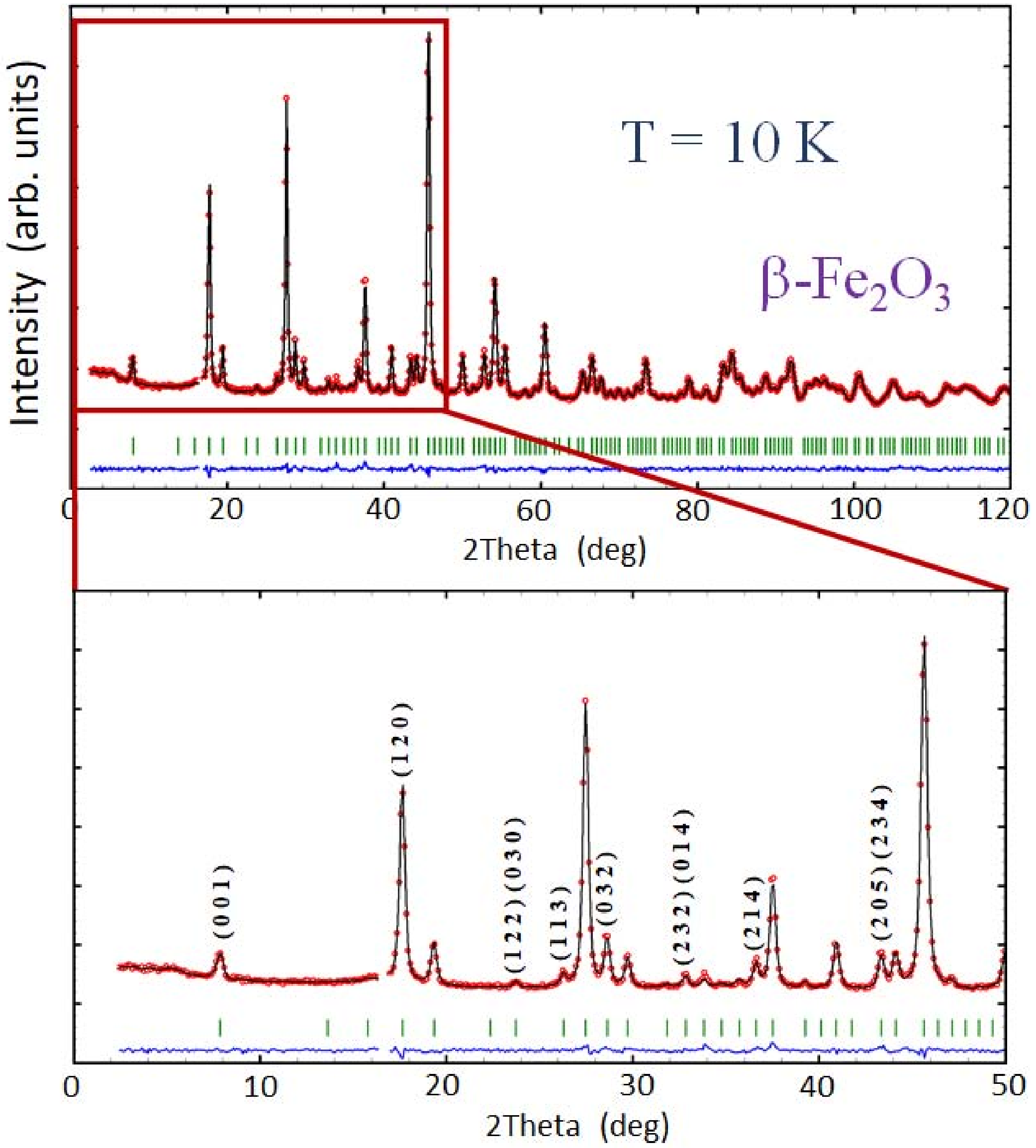

**Figure 6.** Rietveld refinement (black line) of the neutron diffraction pattern measured at 10 K (red circles). An expanded view of the low-*Q* region is also shown, highlighting the main magnetic reflections associated with the magnetic ordering. The 10 K pattern was refined using the primitive cubic $P_Ia$-$3$ symmetry and the magnetic propagation vector **k**=(1,1,1).

As reference magnetic atoms we can consider Fe1 (*8b*) at (0,0,0|$m_x$,$m_x$,$m_x$) and Fe2 (*24d*) at (x,0,1/4|0,$m_y$,$m_z$). The site symmetries of the two positions occupied by magnetic Fe are different. In the case of Fe1, the -3 site symmetry constrains the moments to lie perpendicular to {111} planes, i.e., parallel to the equivalent [111] directions (unit cell diagonals), and |$m_x$|=|$m_y$|=|$m_z$|. The Fe2 atoms have site symmetry 2', thereby forcing the moment component along the 2′ axis to vanish.

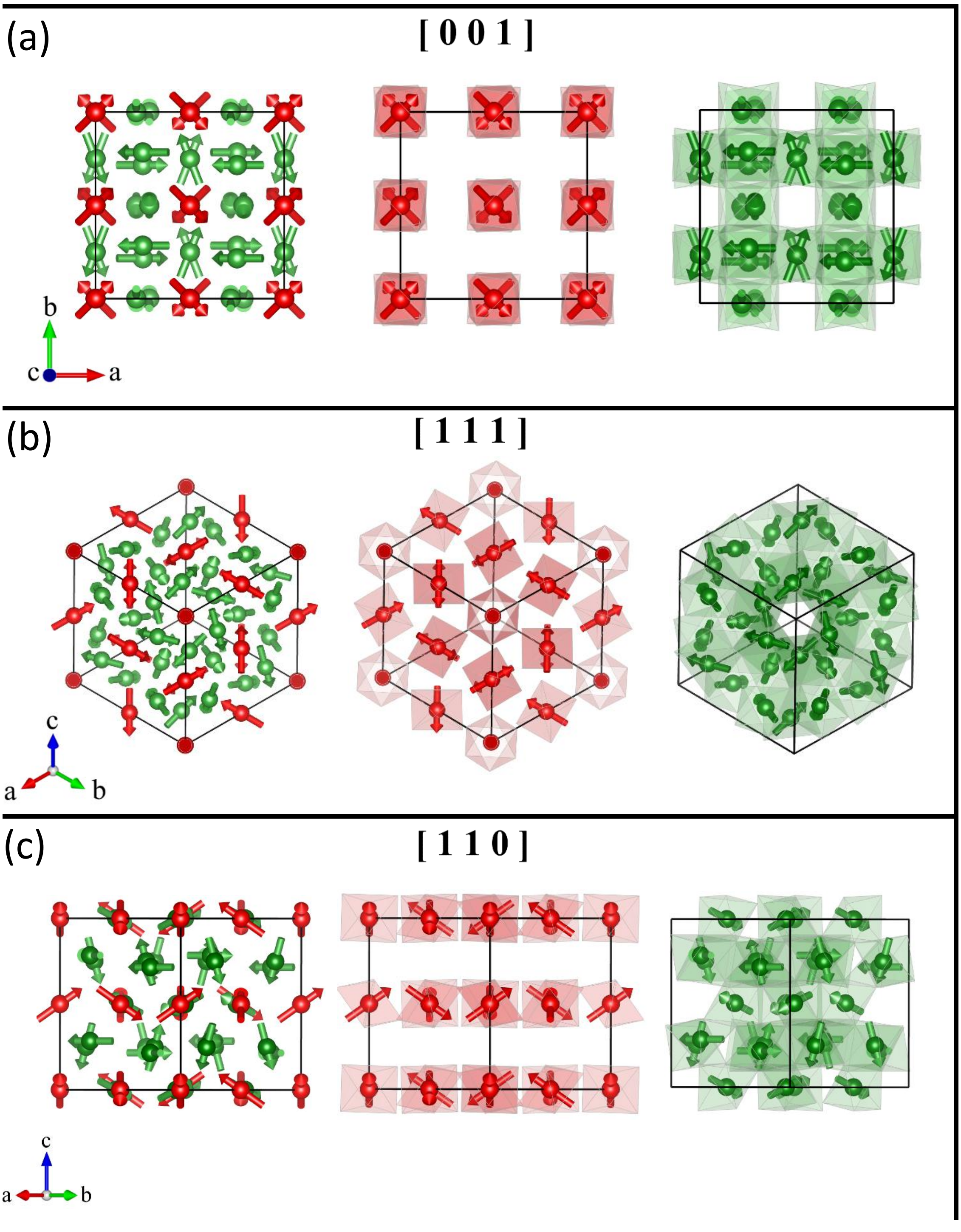

**Figure 7.** Antiferromagnetic order in β-$Fe_2O_3$ (MSG $P_Ia$-3 [No. 205.36], with **k**=(1,1,1)). Several projections of the magnetic structure are provided for clarity: (a) [001], (b) [111], and (c) [110]. The left column shows all magnetic atoms. For clarity, the center and right columns display only the magnetic moments at the Fe1 (red) and Fe2 (green) sites, respectively. Note that the magnetic moments of the Fe1 atoms located at the vertices of the unit cell in panel (b) are pointing out of the plane of the figure.

**Table III.** Magnetic structure of $\beta$-$Fe_2O_3$ (from neutron data recorded at 10 K). Atomic parameters and symmetry operations are given in the standard setting of the MSG ($P_I\,a\text{-}3$).

| **Temperature: 10 K** | **$\beta$-$Fe_2O_3$** | |
|---|---|---|
| Parent space group | *Ia-3* (N. 206) (**a**,**b**,**c**;0,0,0) | |
| Propagation vector(s) | **k** = (1, 1, 1 ) (H-point)<br>magnetic little co-group: *m*-31' | |
| Transformation from parent basis | (**a**,**b**,**c**;0,0,0) | |
| MSG symbol | $P_I\,a\text{-}3$ (No. 205.36) | |
| Transformation from basis used (MSG) to standard setting | (**a**,**b**,**c**;0,0,0) | |
| Magnetic point group | *m-3.1'* (No. 29.2.110) | |
| Unit cell parameters (Å) | *a* =9. 3890 α =90º<br>*b* =9. 3890 β =90º<br>*c* =9. 3890 γ =90º | |
| Generators (symmetry operations) | x,y,z,+1 | { 1 \| 0 } |
| | x+1/2, -y+1/2, -z, +1 | { $2_{100}$ \| 1/2 1/2 0 } |
| | -x, y+1/2, -z+1/2, +1 | { $2_{010}$ \| 0 1/2 1/2 } |
| *Centering:* | -x+1/2, -y, z+1/2, +1 | { $2_{001}$ \| 1/2 0 1/2 } |
| | z, x, y, +1 | { $3^{+}_{111}$ \| 0 } |
| **(1\|0,0,0)** | y, z, x, +1 | { $3^{-}_{111}$ \| 0 } |
| + | -y, z+1/2, -x+1/2, +1 | { $3^{-}_{1-1-1}$ \| 0 1/2 1/2 } |
| **(1'\|1/2,1/2,1/2)** | -z+1/2, -x, y+1/2, +1 | { $3^{+}_{1-1-1}$ \| 1/2 0 1/2 } |
| | -y+1/2, -z, x+1/2, +1 | { $3^{-}_{-11-1}$ \| 1/2 0 1/2 } |
| | z+1/2, -x+1/2, -y, +1 | { $3^{+}_{-11-1}$ \| 1/2 1/2 0 } |
| | y+1/2, -z+1/2, -x, +1 | { $3^{-}_{-1-11}$ \| 1/2 1/2 0 } |
| | -z, x+1/2, -y+1/2, +1 | { $3^{+}_{-1-11}$ \| 0 1/2 1/2 } |
| | -x, -y, -z, +1 | { -1 \| 0 } |
| | -x+1/2, y+1/2, z, +1 | { $m_{100}$ \| 1/2 1/2 0 } |
| | x, -y+1/2, z+1/2, +1 | { $m_{010}$ \| 0 1/2 1/2 } |
| | x+1/2, y, -z+1/2, +1 | { $m_{001}$ \| 1/2 0 1/2 } |
| | -z, -x, -y, +1 | { $-3^{+}_{111}$ \| 0 } |
| | -y, -z, -x, +1 | { $-3^{-}_{111}$ \| 0 } |
| | y, -z+1/2, x+1/2, +1 | { $-3^{-}_{1-1-1}$ \| 0 1/2 1/2 } |
| | z+1/2, x, -y+1/2, +1 | { $-3^{+}_{1-1-1}$ \| 1/2 0 1/2 } |
| | y+1/2, z, -x+1/2, +1 | { $-3^{-}_{-11-1}$ \| 1/2 0 1/2 } |
| | -z+1/2, x+1/2, y, +1 | { $-3^{+}_{-11-1}$ \| 1/2 1/2 0 } |
| | -y+1/2, z+1/2, x, +1 | { $-3^{-}_{-1-11}$ \| 1/2 1/2 0 } |
| | z, -x+1/2, y+1/2, +1 | { $-3^{+}_{-1-11}$ \| 0 1/2 1/2 } |
| Positions of magnetic atoms | Fe1 0.00000 0.00000 0.00000<br>Fe2 0.28410 0.00000 0.25000 | |
| Positions of non-magnetic atoms | O1 0.1429 0. 1315 0.9062 | |
| Magnetic moments components ($\mu_B$), their symmetry constraints and moment magnitudes. | Fe1 2.32(7) 2.32(7) 2.32(7) (mx,mx,mx) 4.02 (12)<br>Fe2 0 -3.42(7) 1.28(5) (0,my,mz) 3.65 (6) | |
| Agreement factors | **$\chi^2$ :** 8.89 **$R_f$ :** 1.50 **$R_B$ :** 2.14 | |

**Table IV**. Decomposition of the magnetic representation of the gray group *Ia-31'* (206.38) (paramagnetic phase) for H-point **k**=(1, 1, 1) in β-$Fe_2O_3$, and positions occupied by the magnetic atoms in the chosen parent setting.

| *Wyckoff position* | *Decomposition into irreps* |
|---|---|
| 8a: (0,0,0) | mH1+(1) ⊕ mH2+H3+(2) ⊕ 3 mH4+(3) |
| 24d: (x,0,1/4) | 2 mH1+(1) ⊕ 2 mH1-(1) ⊕ 2 mH2+H3+(2) ⊕ 2 mH2-H3-(2) ⊕ 4 mH4+(3) ⊕ 4 mH4-(3) |

### III-d Symmetry analysis of the magnetic structure of β-$Fe_2O_3$

As shown above, the antiferromagnetic structure of β-$Fe_2O_3$ is very different from a collinear spin arrangement with antiparallel alignment. In this section we present a detailed symmetry analysis of the ordered magnetic phase in β-$Fe_2O_3$.

In the paramagnetic *gray* group *Ia-31'* (No. 206.38 in BNS notation [35]), Fe cations occupy the *8a* Wyckoff position (0,0,0) (Fe1), and the *24d* position, (*x*,0,1/4) (Fe2). Oxygen anions occupy the *48e* position, (*x*,*y*,*z*), with only one crystallographically independent O site. For simplicity, the parent cell is described using the standard setting, *Ia-31'* (**a**,**b**,**c**;0,0,0), as given in Table I. The paramagnetic and magnetically ordered unit cells are characterized by identical lattice parameters and therefore the same crystallographic and magnetic cell volumes. However, the magnetic transition in β-$Fe_2O_3$ is driven not by zone-center modes, but by H-point (1,1,1) modes of the parent cubic phase. In Table IV we show the decomposition into irreducible representations (irreps) of the magnetic representation of the gray group *Ia*-31' of paramagnetic β-$Fe_2O_3$ for the H-point, **k**=(1,1,1). The decomposition is given for the two Fe magnetic sites. The dimensions of the irreps of the little group for the Fe1 site are 1 (mH1+), 2 (mH2+H3+), and 3 (mH4+). Additionally, the magnetic representation of the lower-symmetry Fe2 site also includes the irreps mH1-(1), mH2-H3-(2) and mH4-(3).

As inferred from the neutron diffraction data, the magnetic ordering is accounted for by a single magnetic irrep. The two magnetic sites order simultaneously and transform as mH1+, adopting the same isotropy magnetic subgroup $P_I a\text{-}3$, dictated by this irrep. This single active mH1+ magnetic irrep is one-dimensional and allows the magnetic phase transition to be continuous according to Landau theory. Moreover, this irrep contains two possible modes ($B_1$(a) and $B_2$(a)) at the *24d* site (Fe2), but only one ($A_g$(a)) at the *8a* site (Fe1).

**Table V**. Relationship between the magnetic structure of β-$Fe_2O_3$ ($P_I\,a\text{-}3$ (**a, b, c** ; 0, 0, 0)) and its parent paramagnetic phase. For each magnetic mode, the base vector and observed amplitudes Cn (in $\mu_B$, n=1, 2, 3) for each site are indicated.

<table>
<tr><td>Compound: β-Fe<sub></sub>$_2$O$_3$</td><td colspan="2">MSG: $P_I\,a\text{-}3$</td></tr>
<tr><td>Parent space group</td><td colspan="2">Ia-3 (N. 206) (a,b,c;0,0,0)</td></tr>
<tr><td>Transformation from parent basis to the one used for the magnetic structure</td><td colspan="2">(a,b,c;0,0,0)</td></tr>
<tr><td>Propagation vector</td><td colspan="2">K = (1,1,1)</td></tr>
<tr><td>Primary irrep(s) label(s ) with dimension</td><td colspan="2">mH1+ (1)</td></tr>
<tr><td rowspan="2">Description of the primary irrep(s)</td><td colspan="2">mH1+:</td></tr>
<tr><td>$\{1\vert 0,0,0\}$ : 1<br>$\{3^{+}_{111}\vert 0,0,0\}$ : 1<br>$\{3^{-}_{111}\vert 0,0,0\}$ : 1<br>$\{-3^{+}_{111}\vert 0,0,0\}$ : 1<br>$\{-3^{-}_{111}\vert 0,0,0\}$ : 1<br><br>$\{-1\vert 0,0,0\}$ : 1<br><br>$\{2_{001}\vert 0,1/2,0\}$ : -1<br>$\{2_{010}\vert 1/2,0,0\}$ : -1<br>$\{2_{100}\vert 0,0,1/2\}$ : -1<br><br>$\{3^{+}_{-11-1}\vert 0,0,1/2\}$ : -1<br>$\{3^{+}_{1-1-1}\vert 0,1/2,0\}$ : -1<br>$\{3^{+}_{-1-11}\vert 1/2,0,0\}$ : -1<br>$\{3^{-}_{1-1-1}\vert 1/2,0,0\}$ : -1<br>$\{3^{-}_{-1-11}\vert 0,0,1/2\}$ : -1<br>$\{3^{-}_{-11-1}\vert 0,1/2,0\}$ : -1</td><td>$\{1'\vert 0,0,0\}$ : -1<br>$\{1\vert 1/2,1/2,1/2\}$ : -1<br>$\{1\vert -1/2,1/2,1/2\}$ : -1<br>$\{1\vert 1/2,-1/2,1/2\}$ : -1<br>$\{1\vert 1/2,1/2,-1/2\}$ : -1<br><br><br><br>$\{m_{001}\vert 0,1/2,0\}$ : -1<br>$\{m_{010}\vert 1/2,0,0\}$ : -1<br>$\{m_{100}\vert 0,0,1/2\}$ : -1<br><br>$\{-3^{+}_{-11-1}\vert 0,0,1/2\}$ : -1<br>$\{-3^{+}_{1-1-1}\vert 0,1/2,0\}$ : -1<br>$\{-3^{+}_{-1-11}\vert 1/2,0,0\}$ : -1<br>$\{-3^{-}_{1-1-1}\vert 1/2,0,0\}$ : -1<br>$\{-3^{-}_{-1-11}\vert 0,0,1/2\}$ : -1<br>$\{-3^{-}_{-11-1}\vert 0,1/2,0\}$ : -1</td></tr>
<tr><td>Description of primary mode(s) and amplitude(s).<br><br>Definition of the mode(s).<br><br>Ia-3[1,1,1]mH1+(a)</td><td colspan="2">mH1+(a), mode 1: [Fe1:a]Ag(a)<br>Fe1 (a) $C_1$=4.006<br><br>mH1+(a), mode 2: [Fe2:d]$B_1$(a)<br>Fe2 (a) $C_2$=1.290<br><br>mH1+(a), mode 3: [Fe2:d]$B_2$(a)<br>Fe2 (a) $C_3$=3.385</td></tr>
<tr><td>Secondary irrep(s) label(s)</td><td colspan="2">Not allowed</td></tr>
</table>

Table V summarizes the information on the active irrep and magnetic modes responsible for the antiferromagnetic order in β-$Fe_2O_3$. The spin arrangement arises from the activation of three modes (of type Ia-3[1,1,1]mH1+(a)) belonging to irrep mH1+. The first mode ($A_g$(a)) orients the Fe1 spins along the cubic crystallographic axes. Its refined amplitude at base temperature is 4.0(1) $\mu_B$. In Fig. 7, the Fe1 moments at the cell corners and at the edge centers are aligned along different ⟨111⟩ cubic axes. The second and third magnetic modes correspond to two perpendicular ordered components at the Fe2 sites, with refined amplitudes of 1.28(5) and 3.42(7) $\mu_B$ at 10 K, respectively. The Fe2 moments are also noncollinear and are not confined to any particular plane. In Fig. 7, the Fe2 B1(a) mode governs the angular deviation of the Fe2 moments from the primitive-cell {100} axes (*a*, *b*, or *c*). The magnitude of the ordered Fe2 moments is determined by the Fe2 B2(a) mode (Table V).

### III-e Magnetic frustration in the bixbyite lattice: the case of β-$Fe_2O_3$

Using synchrotron X-ray and neutron powder diffraction, we investigated possible magnetostrictive effects across the magnetic transition. Fig. 8 shows the temperature dependence of the cubic lattice parameter. Upon cooling, a further contraction of the cubic unit cell occurs across the Néel transition.

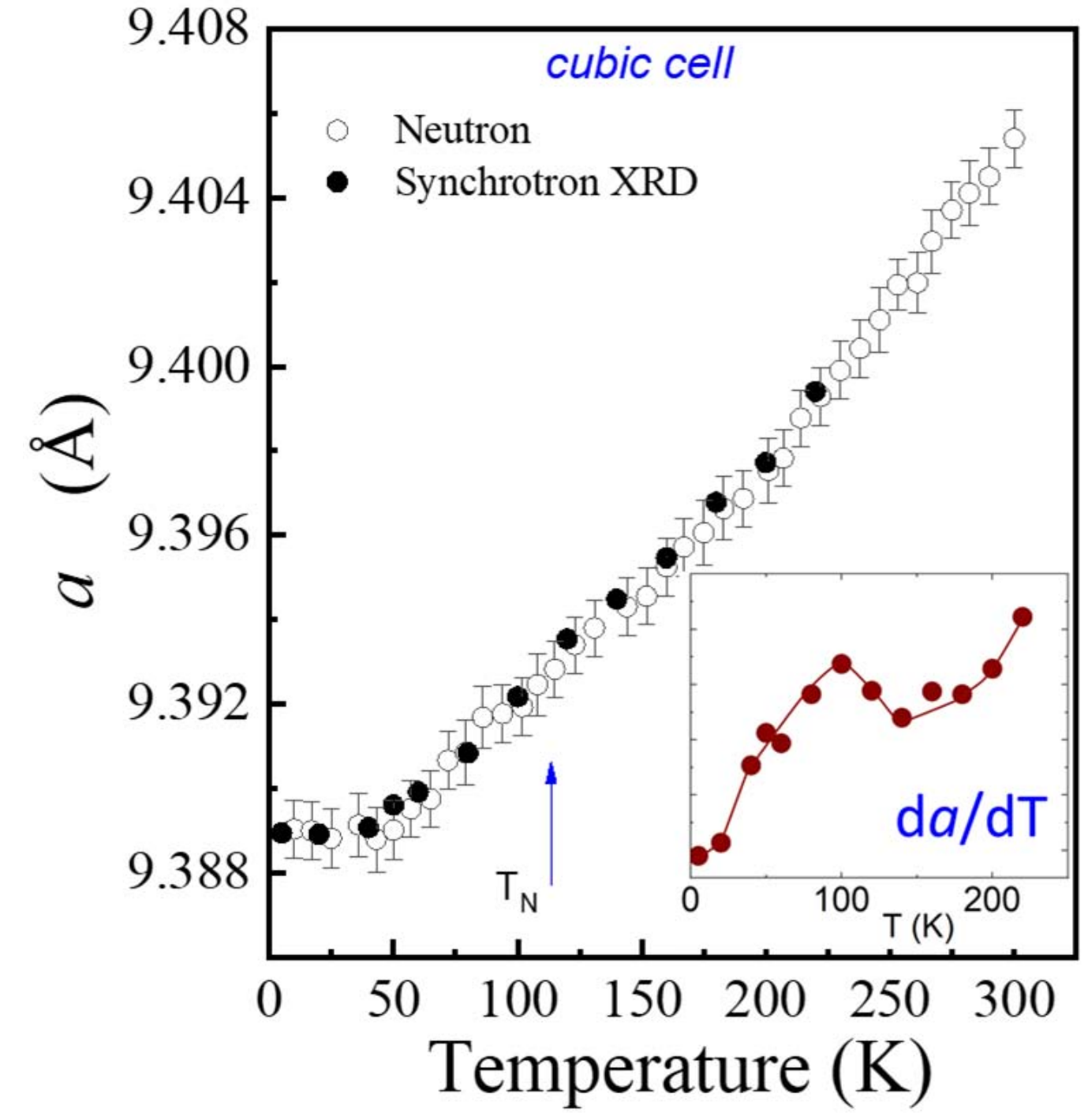

**Figure 8.** Evolution of the unit-cell across the magnetic transition, obtained from synchrotron and neutron diffraction data. Inset: Temperature derivative d*a*/d*T* from synchrotron XRPD data.

This β–$Fe_2O_3$ compound can be considered of Heisenberg type. In a previous Mössbauer investigation of this powder iron oxide, the temperature dependence of the multiplet components provided a critical exponent $\beta$ of the antiferromagnetic phase transition of equal to 0.39(1), in good agreement with the value of $\beta$ = 0.365 expected for a three dimensional Heisenberg magnet [22]. The superexchange interaction between $Fe^{3+}$ cations (governed by Goodenough–Kanamori rules [37,38]) is known to be the main exchange interaction between the half-filled $d^5$ orbitals of $Fe^{3+}$. It is antiferromagnetic for both the 90° and 180° Fe–O–Fe bond angles. Additionally, a contribution from direct Fe–Fe exchange is expected for bond angles ($\varphi$) approaching $\varphi$ = 90º [39]). From Table II, it is clearly seen that most of the $Fe^{3+}$–O–$Fe^{3+}$ bonds in this cubic structure display bond angles ($\varphi$) very close to 100 º, as well as some around 126 º. The inter-sublattice (Fe1-Fe2) and intra-sublattice (Fe1-Fe1, Fe2-Fe2) exchange integrals were estimated according to the formula $J_{ij}(\varphi) = J_{90}\sin^2\varphi + J_{180}\cos^2\varphi$ [39–41], where the parameters $J_{90}$ and $J_{180}$ depend on the specific type of bonding coordination. For octahedral sublattices $J_{90}$=0.75 meV and $J_{180}$=14.0 meV [41].

**Table VI**. Calculated Fe-O-Fe exchange integrals $J_{ij}$ (single path), resultant exchange (Jn) between Fe–Fe pairs, and number (Zn) of neighboring *j* atoms coupled through Jn to a reference atom *i* in β-$Fe_2O_3$ at 300 K.

| Coupling | Exchange path | *i* | *j* | Angle (*deg.*) | $J_{ij}$ (*meV*) | **Jn** (*meV*) | *Z*n | ZnJn (*meV*) |
|---|---|---|---|---|---|---|---|---|
| J1 | $J_1$ | Fe1 | Fe2 | 126.41 | 5.41 | **5.41** | 6 | 32.46 |
| J2 | $J_2$ | Fe1 | Fe2 | 97.98 | 1.00 | **2.10** | 6 | 12.60 |
| | $J_2$’ | Fe1 | Fe2 | 99.37 | 1.10 | | | |
| J3 | $J_3$ | Fe2 | Fe2 | 98.17 | 1.02 | **2.24** | 4 | 8.96 |
| | $J_3$’ | Fe2 | Fe2 | 100.91 | 1.22 | | | |
| J4 | $J_4$ | Fe2 | Fe2 | 126.41 | 5.41 | **5.41** | 4 | 21.64 |

The calculated exchange integrals are summarized in Table VI for individual Fe-O-Fe exchange pathways ($J_{ij}$) and final Fe–Fe exchange pairs (Jn), which integrates the different exchange pathways between two given magnetic positions (one pair of Fe sites). Zn is the number of Jn interactions connecting a given Fe atom (reference atom *i*) to different neighboring Fe2 atoms. As shown in Table VI, J1 and J2 are inter-sublattice couplings (Fe1-Fe2) and J3 and J4 are intra-sublattice couplings (Fe2-Fe2). Notice that all of them are antiferromagnetic. There is no Fe1-O-Fe1 superexchange paths in the structure (weaker Fe-O-O-Fe super-superexchange is not considered). In Table VI, J2 and J3 each represent the sum of two equivalent exchange pathways involving the two oxygen atoms of a shared edge. By contrast, J1 and J4 correspond to exchange interactions mediated by a single pathway through a shared octahedral corner. As seen in Table VI, J1 and J4 are expected to be similar and represent the strongest couplings in the system, since they involve the most open bond angles (~126º) among all exchange pathways in the structure.

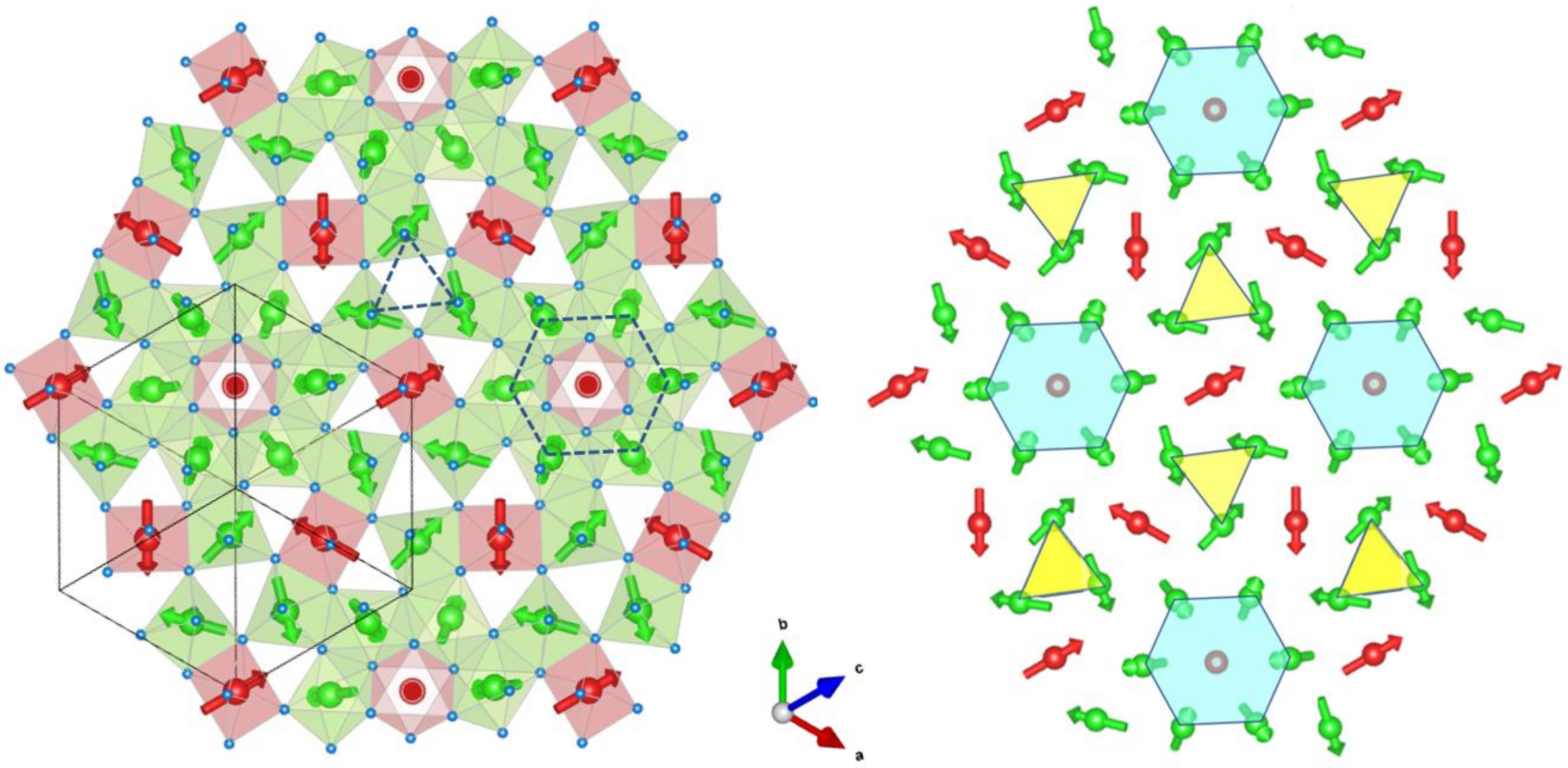

**Figure 9.** Magnetic order in the layer of Fe1 and Fe2 octahedra perpendicular to [1 1 -1]. In this plane, distorted $Fe2O_6$ octahedra (green) form a network of hexagonal rings interconnected by triangular units. Each Fe2 atom is part of a hexagonal ring oriented perpendicular to one of the ⟨111⟩ directions. Likewise, each Fe1 atom is located at the center of one hexagonal ring.

The planes defined by the Fe1 and Fe2 sites perpendicular to the ⟨111⟩ directions are of particular interest for understanding both the magnetic configuration and the origin of geometrical frustration in this structure. Fig. 9 shows a single layer perpendicular to the [1 1 1] direction (in contrast to Fig. 7(b) (left), only one layer of Fe1/Fe2 atoms is shown here). The spin arrangement is shown together with the two types of octahedra associated with the Fe sites. Within the layer, the Fe2 cations form a network that can be viewed as a tiling of two types of interconnected structural units: hexagonal rings and triangles. Each hexagonal ring is formed by six Fe2 cations, and it is surrounded by six triangles, each consisting of three Fe2 octahedra. For clarity, these regular hexagonal and triangular motifs are highlighted in Fig. 9.

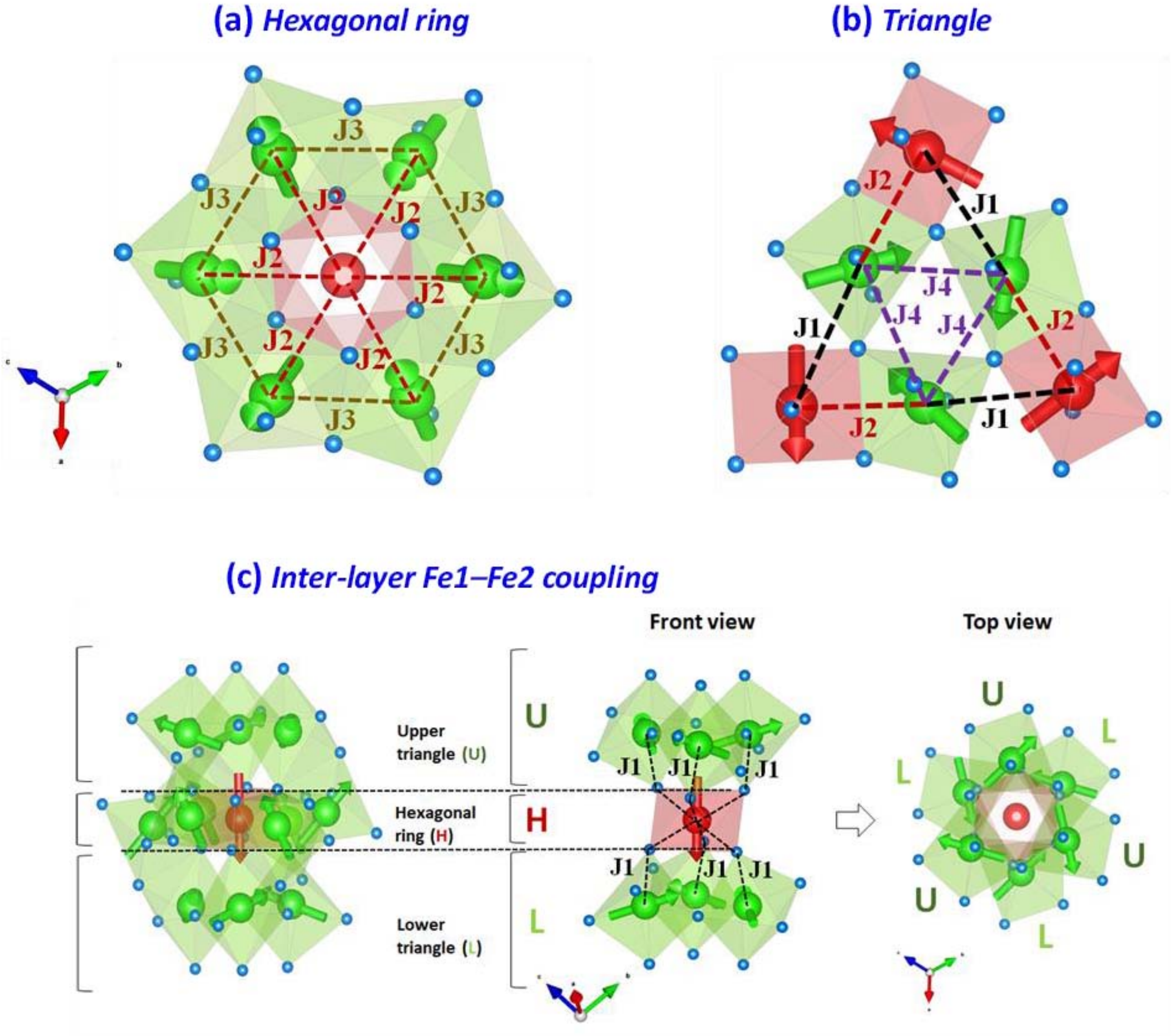

**Figure 10.** Magnetic order and main exchange interactions. Intra-layer: (a) the Fe2 hexagonal rings; (b) the Fe2 triangles. Inter-layer: (c) Fe1–Fe2 bonds between successive [111] layers. (left) Front view. (center) Front view showing Fe1-O-Fe2 exchange bonds (for clarity, the in-plane Fe2 hexagonal ring has been omitted). Fe1, Fe2 and O atoms are shown in red, green and blue colors, respectively. The exchange interactions are listed in Table VI.

Fig. 10(a) shows the spin arrangement in a hexagonal ring. The antiferromagnetic (AFM) exchange interaction between the central Fe1 spin and the six surrounding Fe2 spins occurs via the J2 interaction (~2.10 meV, Fe1–Fe2,) indicated in the figure. Relative to the central Fe1 spin, the Fe2 ordered magnetic moment can be decomposed into two components: (i) an out-of-plane component, identical for all six Fe2 octahedra and antiparallel to the Fe1 moment, and (ii) an in-plane component, oriented radially with respect to the center of the hexagon and with alternating orientation among the six neighboring Fe2 sites that form the hexagon. J2 is responsible for the out-of-plane AFM coupling between the Fe1 and Fe2 moments, whereas J3 determines the in-plane radial AFM arrangement of adjacent Fe2 spins around the hexagon.

With regard to the triangles, Fig. 10(b) presents a schematic representation showing the AFM J4 exchange interaction between the Fe2 cations of the triangle, with their spins 120° apart from each other. Each Fe2 triangle is inscribed in a larger triangle with Fe1 atoms at the vertices, so that each vertex of the Fe2 triangle shares an edge with one Fe1 octahedron (J2) and a corner with another (J1). As indicated in Table VI, the dominant interactions J1 and J4, followed by the weaker J2 and J3, play a key role in stabilizing the spin configuration shown in Fig. 10. It is also worth recalling, in light of Fig. 10, that the -3 site symmetry constrains the Fe1 magnetic moments to lie perpendicular to {111} planes.

To gain further insight into the magnetic order, Fig. 10(c) depicts the interlayer coupling between Fe1 and Fe2 moments in adjacent [111] layers. In the central plane Fe1 occupies the center of a hexagonal Fe2 ring (left panel of Fig. 10(c)). However, along the ⟨111⟩ direction (out-of-plane) the Fe1 atom lies between two triangles formed by Fe2 atoms (center panel of Fig. 10(c)). Explicitly, it lies midway between a triangle of Fe2 octahedra in the adjacent upper layer and another in the lower layer, the two triangles being rotated by 180° with respect to each other. The relative orientation of the ordered magnetic moments is shown (although, for clarity, the in-plane Fe2 hexagonal ring has been omitted in the center and right panels of the Fig. 10(c)). Fe2 moments in the two triangles are canted towards the [1 1 1] cubic axis in opposite sense to the central Fe1 spin (imposed by J1). At the same time, the frustration associated with the AFM J4 interaction between Fe2 spins within each triangle is relieved by the adoption of an in-plane magnetic component exhibiting a 120° twist between neighboring vertices in the triangle.

In this compact structure, the coupling between successive [111] layers also involves Fe2–Fe2 pairs located in triangular and hexagonal units, respectively (see, e.g., Fig. S3 [31]). Furthermore, in Fig. 11 we identify the Fe1 positions located at the center of a hexagonal ring oriented out of the [111] plane. Hence, those Fe1 sites whose easy axis is not perpendicular to the plane occupy the center of another out-of-plane-oriented hexagonal ring, parallel to another of the unit cell diagonals, whose projections onto the [111] plane are represented in Fig. 11 as elongated blue rectangles.

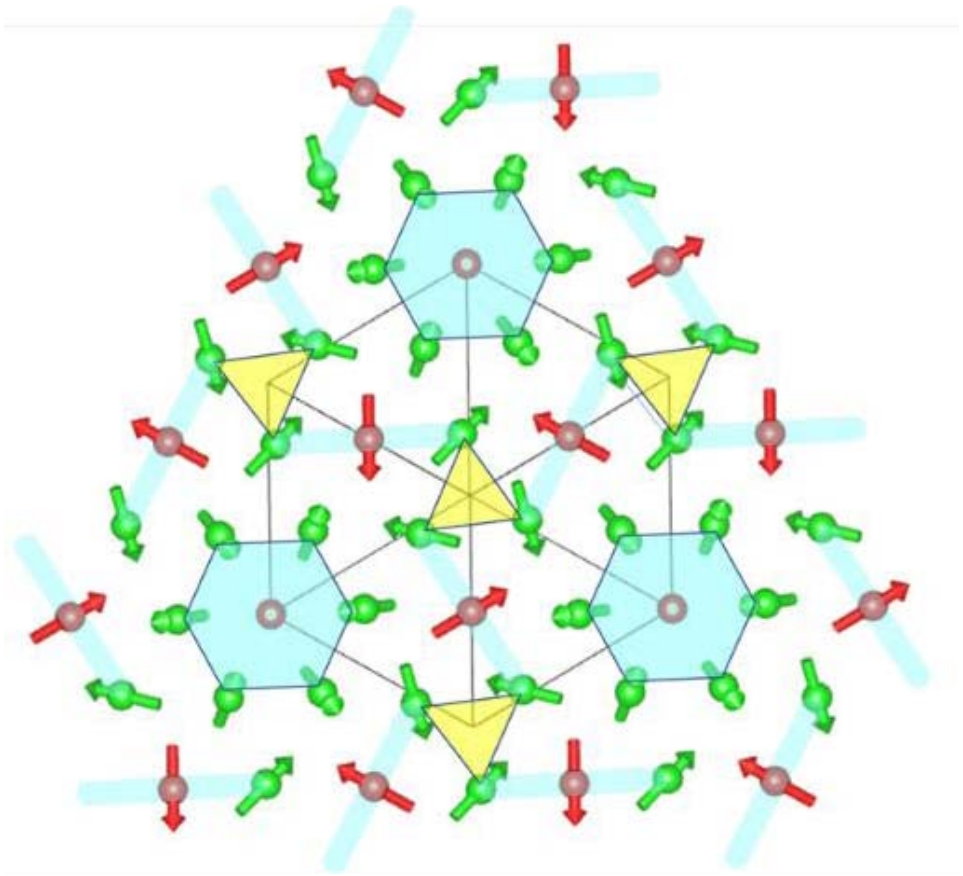

**Figure 11.** Magnetic order in the [111] layer of Fe1 and Fe2 octahedra. Fe2 hexagonal rings and triangular units are highlighted in light blue and yellow, respectively. The projections of Fe2 hexagonal rings oriented both parallel to and out of the [111] plane are shown in light blue. Each hexagonal ring hosts an Fe1 atom at its center.

## IV- DISCUSSION AND CONCLUSIONS

Across the different polymorphs of the abundant iron (III) $Fe_2O_3$ oxide, magnetic properties vary markedly despite their identical chemical composition, being largely governed by structural features that shape the exchange topology, magnetocrystalline anisotropy, and electronic structure. In this work, we have investigated the impact of geometrical frustration on the magnetic properties of the bixbyite polymorph of $Fe_2O_3$. So far, its magnetic properties have remained poorly established compared with those of the rest of $Fe_2O_3$ polymorphs, such as $\alpha$-$Fe_2O_3$ (hematite), $\gamma$-$Fe_2O_3$ (maghemite) or $\varepsilon$-$Fe_2O_3$ (epsilon) stable at ambient pressure. In $\beta$-$Fe_2O_3$, all the $Fe^{3+}$–O–$Fe^{3+}$ exchange interactions (both intra- and inter-sublattice) are antiferromagnetic. Long-range antiferromagnetic ordering suddenly develops through the breaking of the body-centering symmetry, driven by the activation of the irrep mH1+ associated to the H-point (propagation vector **k**=(1,1,1)) and the antitranslation (1'|1/2,1/2,1/2). Below the Néel transition, the magnetic cell becomes primitive, while the antitranslation generates two interpenetrating primitive cubic magnetic cells whose magnetic moments are inverted with respect to each other. The two magnetic sites order simultaneously and the antiferromagnetic ground state adopts the cubic magnetic space group $P_Ia\text{-}3$ (No. 205.36). This Type IV *(klassengleiche)* magnetic group is non-polar, and in this case spontaneous electric polarization or ferromagnetism are both forbidden by the magnetic symmetry.

Through the example of $Fe_2O_3$, we have shown that the bixbyite structure provides an attractive platform for the emergence of noncollinear magnetism and complex magnetic phases, owing to its distinctive structural features, extensively analyzed in our study, which naturally give rise to geometrical magnetic frustration. In addition to the network of interconnected triangles, another notable feature of β-$Fe_2O_3$ and related structural derivatives is the coexistence of potentially magnetic sublattices with distinct point symmetries and easy-axis orientations, which strongly favors the formation of noncollinear magnetic order. A particularly interesting feature of this structure is the presence of hexagonal rings of Fe2 atoms that enclose an Fe1 atom at their center. The Fe1 site shows Ising-type local anisotropy and could therefore function as a switching lever for the magnetic behavior of the surrounding rings.

Within the bixbyite structure, the compound can be formulated as $M1_{0.5}M2_{1.5}O_3$, with M1 occupying the 8*a* site and M2 the 24*d* site. Here, M1= Fe1 and M2= Fe2. Notably, the bixbyite structure is not unique to β-$Fe_2O_3$, but is adopted by a broad range of sesquioxides, a family that remains largely unexplored. As example, structural studies have established the occurrence of bixbyite-type phases in $Mn_2O_3$ and Fe-substituted manganites such as $(Mn,Fe)_2O_3$ [42,43], while studies of $Sc_2O_3$-based compounds have also demonstrated the incorporation of magnetic Fe, Mn and V cations into the $(Sc,M)_2O_3$ bixbyite lattice [44,45], offering an additional route for tuning magnetic interactions through chemical substitutions. In addition, the bixbyite family also encompasses rare-earth sesquioxides, whose stability increases toward the heavier rare-earth elements and mixed rare-earth/transition-metal compounds [46–52]. These findings establish these structural motifs and their associated magnetic frustration as a promising platform for functional materials design, motivating the exploration of alternative compositions, magnetic species with competing exchange interactions, competing anisotropies, and external-field control.

## ACKNOWLEDGMENTS

We acknowledge financial support from the Spanish Ministerio de Ciencia, Innovación y Universidades (MINCIU), through Projects No. PID2021-124734OB-C22 and PID2024-159457OB-C22, cofunded by ERDF from EU, and "Severo Ochoa" Programme for Centres of Excellence in R&D (Grant No. CEX2023-001263-S). Ch.T. was financially supported by China Scholarship Council. Ch.T. work was done as a part of the Ph.D program in Materials Science at Universitat Autònoma de Barcelona. O. M. acknowledges the financial support of the European Union under REFRESH – Research Excellence for Region Sustainability and High-tech Industries (project no. CZ.10.03.01/00/22-003/0000048) and the Research

Infrastructure NanoEnviCz, supported by the Ministry of Education, Youth, and Sports of the Czech Republic (project no. LM2023066). We acknowledge the Institut Laue-Langevin (ILL), the Spanish CRG access program and ALBA the provision of beam time. I. Puente-Orench is acknowledged for technical assistance during neutron measurements.

**Supporting Information description.** Synchrotron XRPD pattern refined at 5 K; difference between the neutron diffraction patterns in the ordered and paramagnetic states; exchange interactions between Fe2 moments in successive [111] layers;

# Supplemental Material

## The bixbyite framework as a platform for frustrated noncollinear magnetism: resolving the magnetic ground state of β-$Fe_2O_3$

Chenjun Tang[1], Ondřej Malina [2,3], Jiří Tuček [4], Francois Fauth[5], Martí Gich[1], and José Luis García-Muñoz[1]*

[1] *Institut de Ciència de Materials de Barcelona (ICMAB-CSIC), Carrer dels Til.lers, 08193 Cerdanyola del Vallès, Spain.*

[2] *Regional Centre of Advanced Technologies and Materials, Czech Advanced Technology and Research Institute, Palacky University, Olomouc, Slechtitelu 27, 77900 Olomouc, Czech Republic.*

[3] *Nanotechnology Centre, Centre for Energy and Environmental Technologies, VSB-Technical University of Ostrava, 17. listopadu 2172/15, 70800, Ostrava-Poruba, Czech Republic*

[4] *Research Centre, Faculty of Electrical Engineering and Informatics, University of Pardubice, Studentská 95 53002, Pardubice, Czech Republic*

[5] *CELLS-ALBA Synchrotron, 08290 Cerdanyola del Vallès, Barcelona, Spain*

* Corresponding Author. Email: garcia.munoz@icmab.es

Content of the *Supplemental Material*:

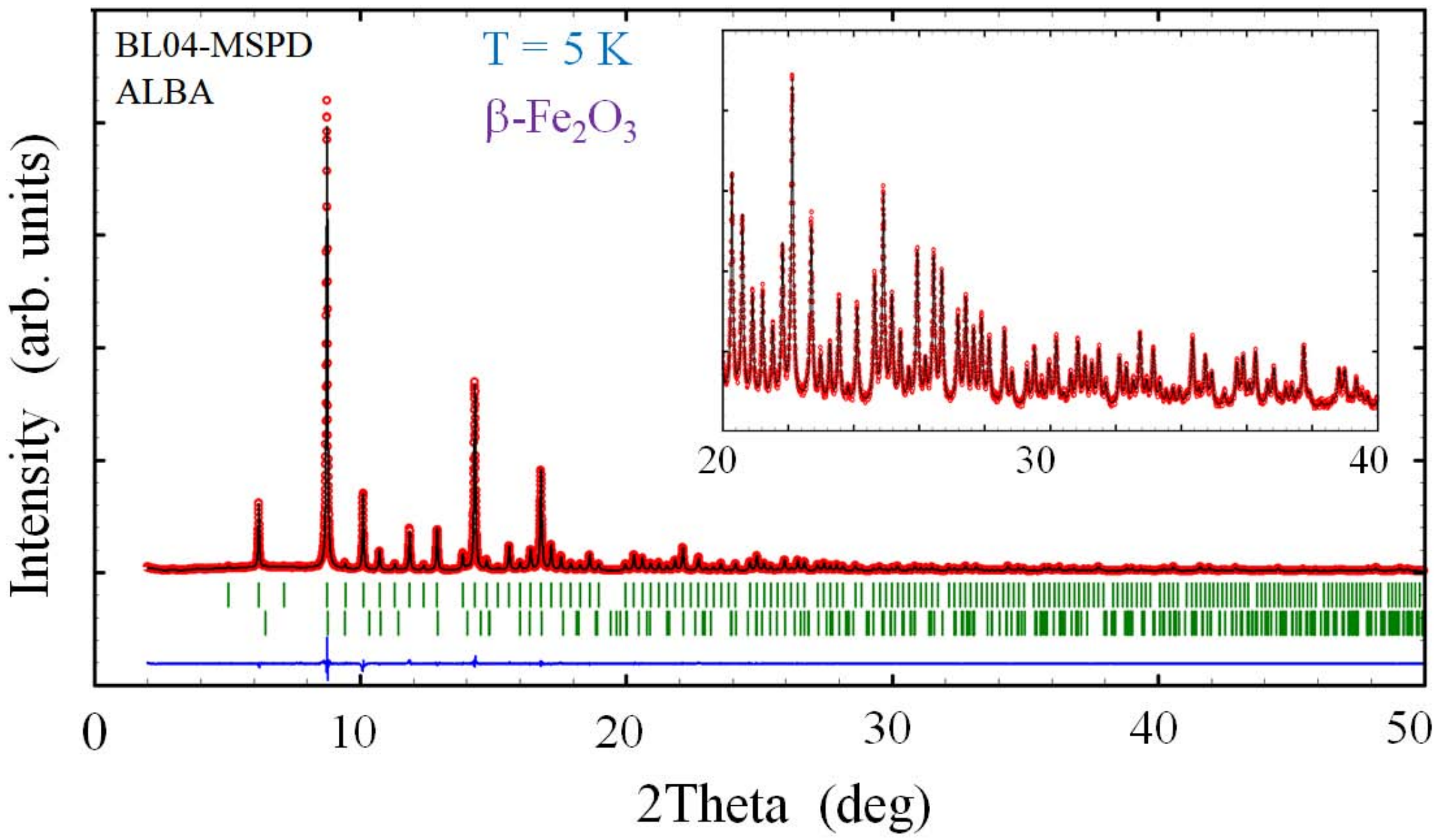

**Figure S1.** Rietveld refinement (solid line) of the synchrotron XRPD pattern (points) for β-$Fe_2O_3$ at 5 K (measured with a high-angular-resolution multianalyzer detector), using the *Ia-3* space group (the fit includes a 1.8% of α-$Fe_2O_3$). Inset: Detail of the fit at high angles.

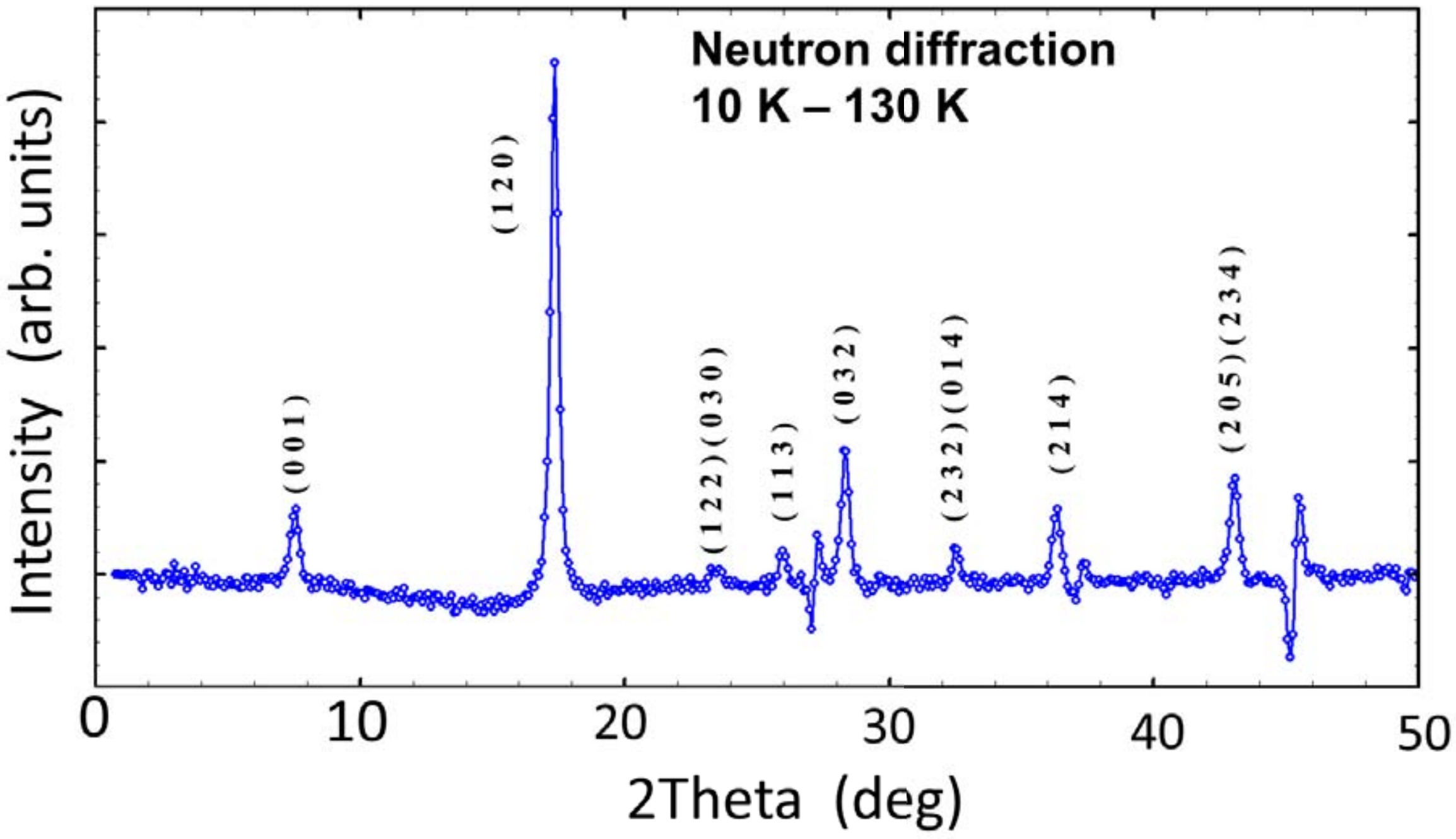

**Figure S2.** β-$Fe_2O_3$. Difference between the neutron diffraction patterns collected at 10 K and 130 K (10 K-130 K). The main magnetic reflections are highlighted (D1b@ILL, $\lambda$ = 1.28 Å). Indexed in the MSG $P_I a\text{-}3$ (205.36).

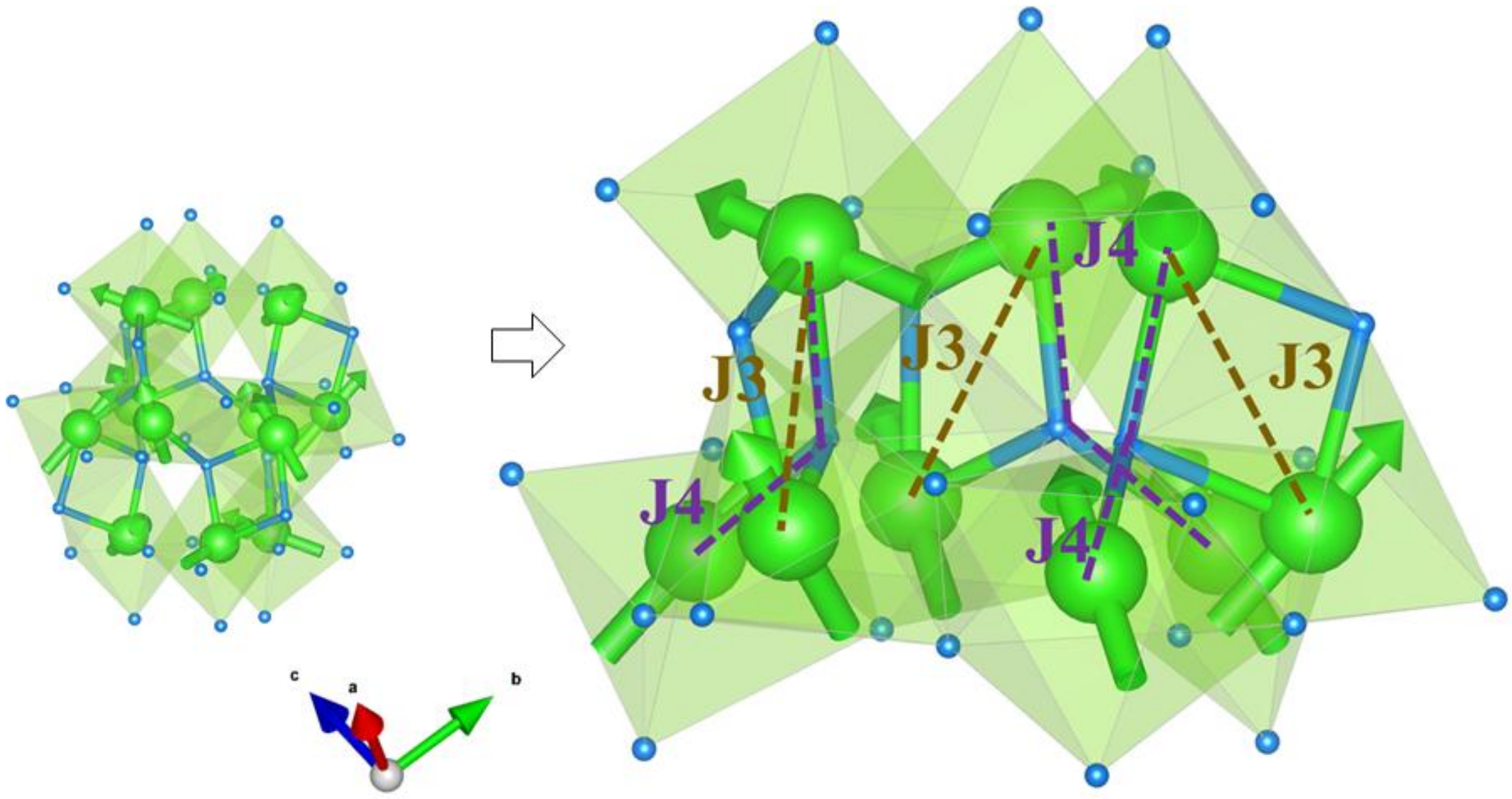

**Figure S3.** Fe2-O-Fe2 exchange interactions between Fe2 moments in successive [1 1 1] layers in β-$Fe_2O_3$.